# Experimental study of the electric dipole strength in the even Mo nuclei and its deformation dependence.

M. Erhard[+], A.R. Junghans, C. Nair, R. Schwengner, R. Beyer, J. Klug[*], K. Kosev, A. Wagner,
Institute of Radiation Physics, Forschungszentrum Dresden-Rossendorf,
01314 Dresden, Germany

E. Grosse, Institute of Radiation Physics, Forschungszentrum Dresden-Rossendorf and
Institute of Nuclear and Particle Physics, Technische Universität Dresden,
01062 Dresden, Germany

**Abstract**

Two methods based on bremsstrahlung were applied to the stable even Mo isotopes for the experimental determination of the photon strength function covering the high excitation energy range above 4 MeV with its increasing level density. Photon scattering was used up to the neutron separation energies $S_n$ and data up to the maximum of the isovector giant resonance (GDR) were obtained by photo-activation. After a proper correction for multi-step processes the observed quasi-continuous spectra of scattered photons show a remarkably good match to the photon strengths derived from nuclear photo effect data obtained previously by neutron detection and corrected in absolute scale using the new activation results. The combined data form an excellent basis to derive a shape dependence of the E1 strength in the even Mo isotopes with increasing deviation from the N = 50 neutron shell, i.e. with the impact of quadrupole deformation and triaxiality. The wide energy coverage of the data allows for a stringent assessment of the dipole sum-rule, and a test of a novel parameterization developed previously which is based upon. This parameterization for the electric dipole strength function in nuclei with A>80 deviates significantly from prescriptions generally used previously. In astrophysical network calculations it may help to quantify the role the p-process plays in the cosmic nucleosynthesis. It also has impact on the accurate analysis of neutron capture data of importance for future nuclear energy systems and waste transmutation.

PACS numbers: 24.30.Cz, 25.20.-x, 26.50.+x, 27.60.+j, 28.90.+i.

[+]) now at INFN, Padova, Italy    *) now at Vattenfall, Stockholm, Sweden

# I. INTRODUCTION: DIPOLE STRENGTH IN HEAVY NUCLEI

The response to electromagnetic radiation plays an important role not only for the fundamental understanding of nuclei but also for the de-excitation processes following nuclear reactions. Details of the dipole strength may significantly affect the de-excitation path after neutron capture in heavier nuclei and it is thus of importance for calculations designed to predict properties of advanced nuclear systems and devices aiming for the transmutation of radioactive nuclear waste. A similar impact is expected on calculations for the cosmic nucleosynthesis, especially for high temperature scenarios where photo-nuclear processes are likely to play an important role. The so-called p-process may be the origin of more than 30 neutron-deficient nuclides not resulting from neutron-capture reactions. They may eventually be produced in the intense photon flux leading to the photo-disintegration of previously formed heavier nuclides [1, 2].

Nuclear photo-effect calculations are usually based on the statistical Hauser-Feshbach theory [2] and these need information on the photon strength functions $f_\lambda$. As the electric dipole mode E1 contributes the most, the photo-disintegration cross sections like ($\gamma$,n) are often used as a measure for the electric dipole strength above the neutron threshold $S_n$. Above the particle-separation energies and also in the GDR region the contribution of ($\gamma,\gamma$) is usually small but below it is dominant. At low energies three features have been discussed for decades [3-5] to be of importance for the physics related to the electric dipole strength function:

(a) The fall-off [4, 5] of the E1-strength on the low-energy slope of the GDR, and near the particle emission thresholds, where it dominates the nuclear photo-effect.

(b) The E1-strength between the states populated by neutron capture and low energy levels, in comparison to nuclear resonance fluorescence (nrf) connecting the ground state to dipole (and quadrupole) excitations. The relation to the regime (a) has attracted special interest [5].

(c) The occurrence of additional 'pygmy' structures [5-8], which show up close to $S_n$. In spite of their relatively low strength as compared to the GDR they may well lead to an enhancement of photo-disintegration processes when $S_n$ gets smaller with increasing distance to the valley of stability.

This possible enhancement has induced the experimental investigation of exotic nuclei with small $S_n$ where the 'pygmy' structures directly enhance the photodisintegration [7]. These data show a structure not quite as broad as the GDR, but wide as compared to the average level distance D.

Many other recent experiments on 'pygmy' resonances concentrated on regions near the magic neutron numbers 82 and 126; the respective data [6, 8] show a number of well separated peaks. This structure was assumed to possibly be of similar nature as the one seen in exotic nuclei and the investigation in dependence of the nuclear charge Z showed an increase of its strength with the N/Z ratio. The experimental techniques applied in that work did not yield information on eventually missing strength distributed over many weak levels hidden in the continuum below the isolated peaks. At variance with that earlier investigations with tagged photons [4, 5, 9] indicated the presence of considerable total photon strength for the range below $S_n$ albeit they only obtained a much lower energy resolution. On the basis of these data an extrapolation of the GDR-Lorentzian, i.e. its tail, was proposed to allow a quantitative prediction of the strength at the particle thresholds. Here the width of the Lorentzian enters as an important parameter as the height of the tail is nearly proportional to it. As long as the total resonance integral is not kept fixed, a fit of the peak region [10] is rather insensitive to the width, as a given cross section maximum can belong to a wide Lorentzian with a large integral or a narrow one with a small total integral. On the other hand the deformation induced splitting of the GDR is well distinguished from its width only in case of well deformed nuclei, which show a clear double humped GDR. For less deformed nuclei this deformation induced splitting may be misinterpreted as a large spreading width with the consequence of a cross section enhancement at the particle thresholds. It is thus important for the determination of Lorentzian parameters to have information on the complete strength also in the tail region. As for most nuclei the excitation region below $S_n$ and $S_p$ is still rather badly known experimentally, a dedicated study of that strength is indicated.

The work described in this paper presents measurements of the electric dipole strength in stable even Mo-isotopes and it covers the low energy tail as well as its continuation up to the GDR maximum. It shows the impact of such a strength determination to statistical Hauser-Feshbach calculations of importance for the understanding of photo-nuclear processes. It aims to completely derive the dipole strength from the data by regarding not only its part seen as discrete lines in nrf-data. By also including the yield showing up as a quasi-continuum below the discrete peaks the full dipole strength is directly determined experimentally in a wide excitation energy range. In that respect the present study is at variance with the previously mentioned [6-8] investigations of 'pygmy' structures which concentrate on discrete spectral lines. The deformation dependence can be well addressed in the Mo-isotopes reaching from the semi-magic $^{92}$Mo to $^{100}$Mo with its triaxially deformed ground state. It will be shown that in these nuclei previous attempts of fitting a Lorentz curve to the peak region of the GDR [10] yield ambiguous extrapolations to lower excitation energies and thus do not allow satisfactory predictions for the strength close to the thresholds.

Previous procedures proposed to derive the E1-strength down to energies near and below $S_n$ have suffered from the following deficiencies:

(a) The measurement of ($\gamma$,n) cross sections by detecting the neutrons suffer from uncertainties in the absolute scale [10-12] and when approaching $S_n$ isotopic target impurities and other background competes especially strong with the decreasing cross section;

(b) if other photo-disintegration channels are open, the photon strength function cannot be deduced from the ($\gamma$,n) cross section alone;

(c) for even nuclei, the study of E1 strength via gamma decay following neutron capture is ambiguous, as only in the rare cases of j=1/2⁻ or 3/2⁻ target ground states s-wave neutron capture populates 1⁻-levels;

(d) experiments performed with He ions [13] populate a diffuse distribution in excitation energy, spin and nuclear orientation quite different from the excitation caused by photons; additionally one can extract their absolute strength values only through using level density information from other sources and a calibration based on other data, *e.g.* from (c);

(e) previous photon-scattering experiments [14-17] using low energy bremsstrahlung delivered strength information only for a limited excitation energy range. At higher photon energies multi-step gamma decays have to be considered explicitly as well as the quasi-continuous part of the spectra mainly resulting from numerous weak resonances.

In earlier publications [18-19] this last point was worked out for $^{98}$Mo and $^{100}$Mo, where the same experimental method as presented here was used. As demonstrated in that work a hitherto unexplored method was found for the data analysis which allows to derive a strength function covering a wide range of excitation. It reaches from just above an energy, where the photo-absorption cross section is exhausted by scattering via isolated levels, up to regions of resonances overlapping within the experimental resolution. For $^{98}$Mo and $^{100}$Mo the strength resulting from this analysis is an obvious continuation of the electromagnetic strength observed as photo-neutron emission [20] above $S_n$. The combined strength function as derived from the two experiments is reasonably well described [19] by using Lorentzians of 4 MeV width; to account for the presumably non-axial deformation up to three components were used. This approach of one width for all GDR-components was quite satisfactory in all even Mo-isotopes [21]. But – as shown recently [22] for well deformed nuclei with A>80 – two-component Lorentzians describe the relative heights of the two GDR-peaks only if the width of each component depends on its pole energy. Here the ground state deformation enters in the calculation of these energies. For surprisingly many nuclei for which respective data have been found in the literature – originating from different experimental methods – also the tails extrapolated from the Lorentzians account well for the E1-strength observed below threshold [22]. There is no need of a width varying with photon energy – at variance with previously given formulae and tabulations [2, 23-24] most of which use an energy dependent width. In this paper we present data for the photon strength covering a wide energy range in the even Mo isotopes as obtained from photon scattering and nuclear photo effect experiments and the results are compared to the parameterization presented in [22], the main ingredients of which will be discussed below. The selection of the stable even Mo isotopes allows to combine information on GDRs and their tails for a rather long isotopic chain. In all the even Mo isotopes ($\gamma$,xn)-reactions have been studied [20] by neutron detection – in view of the large quantities of enriched isotopes needed this is not the case for many other Z.

Mo isotopes have attracted much attention concerning their cosmic synthesis [25] and they are of interest in nuclear technology as part of the fuel matrix, as high melting point material and as steel additive. Of further interest here is the 66 h radio-nuclide $^{99}$Mo, which – if not transmuted before – decays to $^{99}$Tc; whose $2 \cdot 10^5$ year half life makes it one of the long living components of nuclear waste. Special nuclear structure interest in the stable Mo-isotopes arises from the increase of deformation as observed in the spectroscopy of their low lying states; recently a clear experimental signature for triaxiality was found in $^{98}$Mo [26]. To properly assess observations [5-8, 18, 27] on additional "pygmy" structures in photo-absorption data, a proper knowledge of the underlying smooth strength function is of obvious value. It allows to specify how much of the strength found exceeds an underlying 'background strength', *e.g.* the one from a smooth extrapolation of the GDR. The work presented here concentrates on energies above 4 MeV, where photo-absorption reaches predominantly electromagnetic dipole modes. Dipole strength below 4 MeV was studied in detail previously [14-17, 28].

## II. EXPERIMENTAL PROCEDURE

The experiments were performed at the superconducting electron linac of the radiation source

ELBE [29] at Dresden-Rossendorf using a bremsstrahlung continuum. This was either produced by the electron beam hitting a thin Nb-foil or in a solid graphite beam dump. As will be shown below, this parallel use of the e-beam at two sites was very useful for the determination of absolute yields. The set-up has been described (and shown in a top view) previously [12, 30]; a side view is presented in Fig.1. The area directly behind the electron beam dump was used as a site for high flux ( $\sim 10^6$ [s keV cm²]$^{-1}$ at $\sim 7$ MeV) irradiations. In the bremsstrahlung cave targets could be activated simultaneously at two orders of magnitude lower flux, which was well determined employing scattering measurements. Four HPGe-detectors are mounted to observe photons scattered by the Mo-targets and by $^{11}$B added for the flux determination. The Nb radiator thickness of 3.4 mg/cm$^2$ (corresponding to 3.4 10$^{-4}$ radiation lengths) is limited by small angle scattering of the electrons to assure that nearly all of them pass the dump entrance after being magnetically deflected by 45°.

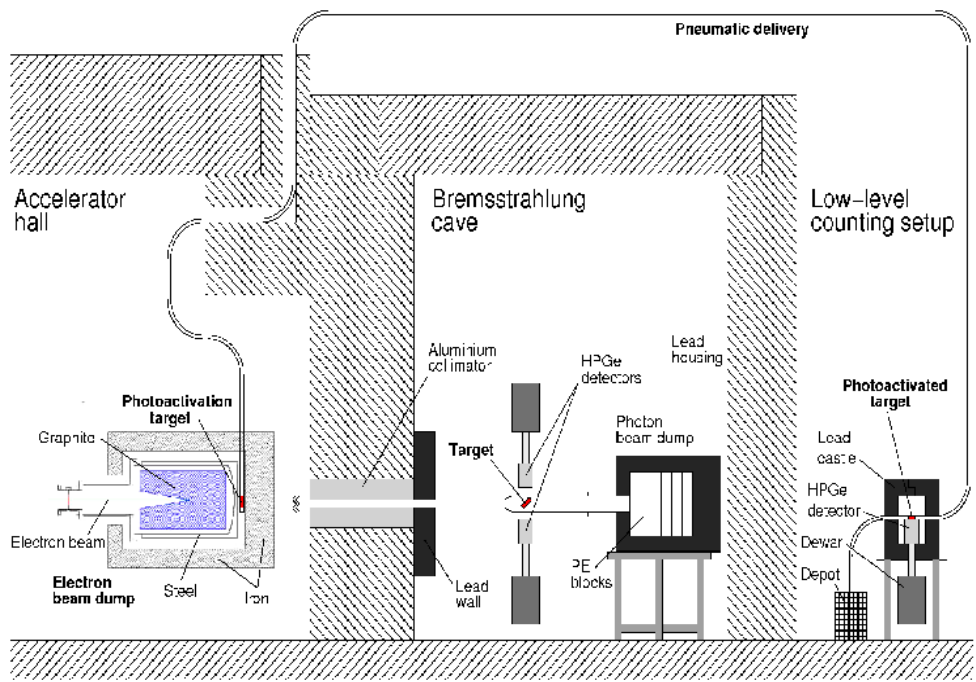

**Fig. 1:** Schematic side view of the bremsstrahlung set-up at the radiation source ELBE. At the left high intensity photons coming from the electron beam dump behind a 45° deflection magnet (not shown) produce photo-activated nuclides to be studied at the right. In the middle part a well collimated photon beam from a thin Nb radiator (not shown) is used to investigate photon scattering processes.

The bremsstrahlung spectrum was calculated by different procedures as demonstrated in Fig.2. The obvious shortcoming of GEANT-3 [35] could be traced back to an erroneous implementation of the bremsstrahlung tables [37]. The GEANT-calculations used in our studies were performed with a code modified to correct for that error. The calculations with the MCNP code [36] are derived from a widely used compilation [37]. A recently performed calculation [38] based on a quantum-mechanical treatment [39] and an atomic screening calculation [40] agrees well with the tables [37]. As shown in the insert only very close to the endpoint a small excess of the tabulated values vs. the full calculation is found. Within the uncertainty of our near-threshold data this difference is of minor importance.

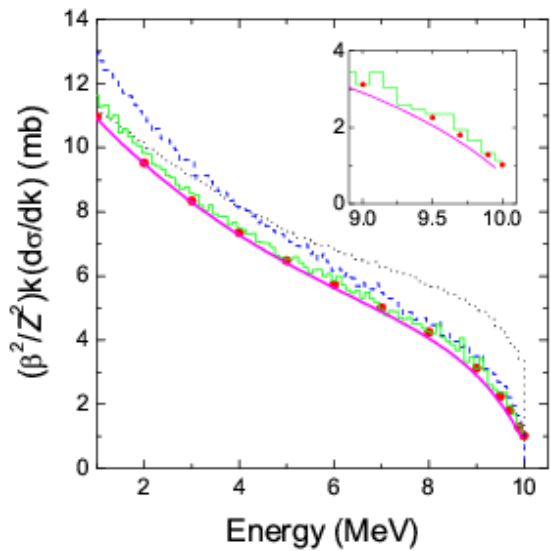

**Fig. 2:** (Color online) Monte Carlo simulations [31] of the bremsstrahlung spectrum produced in a Nb (Z=41) radiator of 3.4 mg/cm$^2$ thickness and an electron kinetic energy of 10 MeV; k is the photon energy and $\beta$ the electron velocity in units of c. The dotted and dashed histograms are calculated with GEANT-3 and GEANT-4 [35], respectively. The full green histogram corresponds to a calculation using MCNP-4C2 [36], which is based on tabulated values [37] (red dots), and the full magenta line depicts the recent bremsstrahlung calculations from ref. [38]. The bremsstrahlung cross sections corresponding to the latter three results are shown in the insert to give an expanded view of the endpoint region.

The knowledge of the bremsstrahlung endpoint energy is important – especially for photonuclear data taken directly above threshold. The electron momentum is not known well enough from the magnetic rigidity in our nearly achromatic beam transport system. This is why we developed an independent method to determine the photon spectrum by measuring the energy of protons from the break-up of deuterium. A $[C^2H_2 \cdot C^2H_2]_n$ - foil of 4 mg/cm² thickness stayed in the photon beam near the exit of the collimator during all measurements. Si-detectors were used to determine the energy of the protons, and proper account was made for energy loss and kinematic broadening [12, 30]. The uncertainty in the endpoint energy was reduced to 50 keV and its effect on the activation yields is included in the error bars in Figs. 8 and 10. In the photon scattering experiment an inaccuracy in the energy determination of 0.1 MeV would lead to an error of only 1 % in the cross section at 9 MeV. As described in previous publications [12, 17, 27, 30] the absolute flux normalization was determined by analyzing the photon scattering from $^{11}$B and using ground state transition strengths determined previously by resonant self-absorption [33, 34]. Because of this use of the high energy transitions in $^{11}$B for the flux determination the final uncertainty is determined by these data; for 7.29 MeV an accuracy of 7 % was reported [34].

By comparing the gamma radiation from Au samples simultaneously activated at both positions a flux normalization for the activation site was determined relative to the scattering site. Different absorption in the two photon paths was determined by MCNP-simulations [36] and correction was made for the non-negligible resonant self-absorption in the $^{11}$B-component of the target [34]. The use of $^{197}$Au as a standard for the photon flux determination in the activation experiments was described earlier [12]. The different targets ($^{11}$B, Mo, $^{197}$Au; their mass was determined by weighing) are hit by the same photon fluence when exposed simultaneously to the photon beam. In all cases the samples were positioned such that they were only exposed to the homogeneous middle section of the photon flux. When we also analyzed the activity induced in the enriched Mo targets during scattering experiments, we obtained consistent results. For the photo-activation measurements many independent samples of natural Mo were used. The samples were irradiated for 8-36 h for the study of long lived activities off-line. HPGe-detectors in low activity lead shields were used in a close gamma-detection geometry and summing effects caused by the small distance from the activated sample have been properly handled. To have access to the 65 s isomer of $^{91}$Mo a pneumatic delivery system and multiple short irradiations were used [12, 31]. This allowed an efficient study of the ($\gamma$,n) reaction to $^{91m}$Mo, which decays to $^{91}$Nb. This path competes with the direct production of $^{91}$Nb via ($\gamma$,p) at energies above $S_n$ in $^{92}$Mo, but below $S_n$ our set-up allows an unperturbed study of $^{92}$Mo ($\gamma$,p). The importance of this information for the normalization of the dipole strength function will be shown below. The main effect of our activation data is that they allow a test of the ($\gamma$,n)-data obtained by directly observing the neutrons. Discrepancies for such measurements have caused intensive dispute [11] and for the isotopes $^{92}$Mo and $^{98}$Mo [20, 32] disagreeing data have been reported.

The photon scattering experiments were performed in the cave area shown in the middle part of Fig.1. An electron kinetic energy of 13.2 MeV (13.9 MeV in the case of $^{92}$Mo) was used to produce bremsstrahlung in a Nb foil of 3.4 mg/cm$^2$ and an Al cylinder of 27 g/cm² thickness reduced the strong low energy component of it. At that electron energy the photon flux passing through the collimator made of pure Al was in the range of several $10^4$ (s keV cm²)$^{-1}$ for $E_\gamma \sim 7$ MeV. The HPGe-detectors of ~100 % efficiency relative to a 3"x 3" NaI scintillator were surrounded by BGO escape suppression shields to veto signals not depositing the full photon energy in the detector. The photons travelling to them from the target were collimated by 60 mm long Pb-collimators fully covering the front of the BGO. The data discussed here were taken at 127° with respect to the beam and low energy photons were reduced in intensity before entering the Ge by 9 g/cm² of Pb and 2.6 g/cm² of Cu. Two detectors placed at 90° had an additional absorber of Pb with 5.6 g/cm². As described previously [19, 27, 30] the spectral background was also reduced by a proper selection of all material eventually hit by scattered photons. An important ingredient of the experiment is the massive beam dump heavily reducing the background radiation caused by the beam passing the target. The data were accumulated in Compton suppressed spectra using analog to digital converters with a range of 16384 channels. Using a gain of ~ 1 keV/channel allowed to store the data with an rms-resolution of < 5 keV which was reached up to 13 MeV. As described for photon scattering at lower energy [14-15], a comparison of the line integrals for the target nucleus under study to the ones belonging to a calibration nucleus (e.g. $^{11}$B) allows to directly determine an absolute ground state transition width – if feeding and branching can be neglected. This is not the case here, and in Fig. 2 of a recently published Rapid Communication [41] on part of the data also presented here, the effect of correcting for feeding and branching is shown to be important.

As reported previously [19, see Fig.2; 27] the efficiency of the photon detectors was determined by a detailed Monte-Carlo-simulation on the basis of an exact knowledge of the detector geometry including the Pb-collimation in front and the BGO-shield around the detectors. For low photon energies the absolute efficiency as well as its energy dependence was found to be in agreement with gamma-ray intensities from radioactive sources of strength known within 3%. The detector response at $E_\gamma$ = 6.17, 10.76 and 12.14 MeV was connected to the low energy region by using well known cascade decays of strong resonances. For that purpose the HPGe-detectors of this study (with escape suppression) were included in (p, $\gamma$)-experiments [42] on $^{14}$N, $^{27}$Al and $^{11}$B at the beam of the FZD-Tandetron. The simulations were found to be reproduced on an absolute scale within 13 %.

The energy spectra of photons scattered from $^{92}$Mo or $^{94}$Mo targets are displayed in Fig. 3 for bremsstrahlung with 13.2 MeV endpoint energy; for $^{96}$Mo 13.9 MeV were used. The spectra are shown before and after the correction for the detector response. A very large number of lines is seen above 4 MeV which are especially pronounced for $^{92}$Mo. The steep decrease of the photon yield for $^{94}$Mo and $^{96}$Mo is due to the ($\gamma$,n)–thresholds; for $^{92}$Mo $S_n$= 12.67 MeV is just inside the plot, but a less pronounced fall off is already seen above $S_p$=7.46 MeV. The enhancement seen above the line depicting the non-resonant scattering reaches from ~ 4 MeV to the particle threshold; it contains most of the information of importance for the present study.

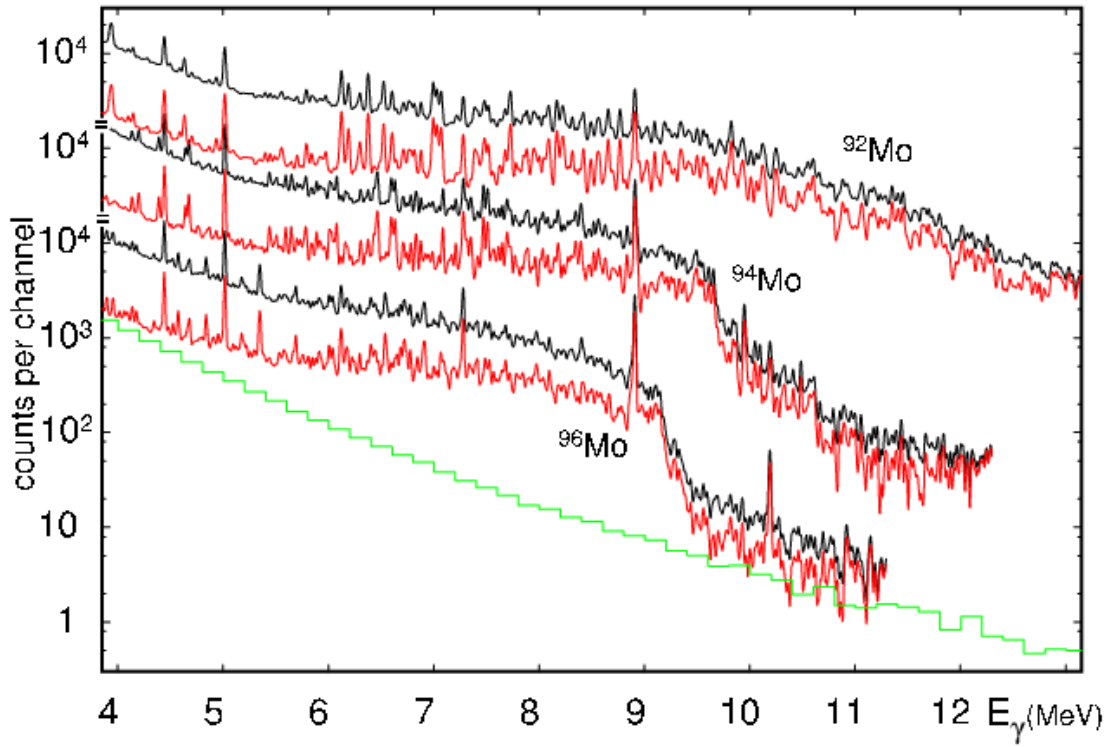

**Fig. 3:** (Color online) Energy calibrated spectra of bremsstrahlung photons scattered from $^{94}$Mo and $^{96}$Mo as produced with an endpoint energy of 13.2 MeV and 13.9 MeV in case of $^{92}$Mo. The strong peaks at 4.44, 5.02, 7.29 and 8.92 MeV result from the $^{11}$B added for the photon flux calibration; the gamma line at 10.2 MeV is probably due to summing of a gamma cascade following neutron capture by $^{73}$Ge. The measured spectra are shown in black, the data resulting from the correction for detector efficiency and de-convolution of the detector response function are shown in red (arbitrarily normalized at 12 MeV to the original spectra). The histogram depicts (for the scale of $^{96}$Mo) the atomic and non-resonant nuclear scattering obtained in a simulation with a modified version of GEANT-3 [19, 35]; it can be used for all three targets which had nearly the same mass.

## III. DATA ANALYSIS
### A. Photon scattering

In nuclear resonance fluorescence (nrf) experiments at low beam energies [14, 15, 17, 28] levels with sufficient transition strength to the ground state are observed as resonances in elastic scattering, and the signal from a scattered photon identifies their excitation energy. At variance with this situation there are two facts strongly influencing photon scattering studies for $E_x > 4$ MeV in heavy nuclei:

(a) Porter-Thomas fluctuations [43] cause a large number of weak transitions with an average level distance similar to the experimental resolution. This leads to a quasi-continuum underlying the stronger peaks, whose heights are also influenced by fluctuation effects.

(b) The depopulation of excited states may lead to multi-step gamma-cascades such that branching ratios have to be regarded as well as possible feeding from above the level.

As will be obvious in the following, point (a) is accounted for in our data analysis by mainly regarding data averaged over 200 keV in the photon energy $E = E_\gamma$; concerning point (b) additional measures have been taken as described in [27] and discussed in the following.

For spin 0 nuclei the dipole strength function $f_1$ is related to the average photon absorption cross section by [4, 5]:

$$f_1(E) \equiv \frac{1}{3(\pi \hbar c)^2 \Delta} \int_\Delta \frac{\sigma_\gamma}{E} dE = \sum_{R=1}^{N} \frac{2 I_R}{3\pi (\pi \hbar c)^2} \frac{E \, \Gamma_{\gamma R}}{(E_R^2 - E^2)^2 + E^2 \Gamma_R^2} \quad (1)$$

where the averaging interval $\Delta$ has to be sufficiently narrow to neglect the variation in $E$. The interval $\Delta$ contains N absorbing levels $R$, which are resonances of ~1 eV width, and $\Gamma_R$ ($\Gamma_{\gamma R}$) are their total (partial) ground state transition widths. This formula for the nuclear resonance absorption cross section is derived in analogy to a classical dipole oscillator with a pole energy $E_R$. The widths are taken to be constant in the region of the resonance as (1) already accounts for the dependence of $\Gamma_{\gamma R}$ upon $E^3$. As resonant absorption also occurs at energies above threshold the total width $\Gamma_R$ may also have contributions from particle emission channels; in general one has:
$$\Gamma_R = \Gamma_{\gamma R} + \Gamma_{nR} + \Gamma_{pR} + \ldots \quad (2)$$

For the Lorentzians as appearing in (1), photo-absorption width and resonance integral $I_R$ of the absorption cross section $\sigma_\gamma$ are related [14]; in even nuclei one has:

$$I_R = \int_{res\, R} \sigma_\gamma\, dE = 3(\pi\hbar c)^2 \frac{\Gamma_{\gamma R}}{E_R^2} \qquad (3)$$

Then $f_1$ can thus be related directly to the partial widths $\Gamma_{rR}$ and their averages $\overline{\Gamma_\gamma}$:

$$f_1(E) \equiv \frac{1}{3(\pi\hbar c)^2 \Delta} \int_\Delta \frac{\sigma_\gamma}{E} dE \approx \frac{1}{3(\pi\hbar c)^2} \frac{1}{E_R \Delta} \int_\Delta \sigma_\gamma\, dE \approx \frac{\sum_R^N \Gamma_{\gamma R}}{\Delta\, E^3} = \frac{\overline{\Gamma_\gamma}}{E^3} \frac{N}{\Delta} = \frac{\overline{\Gamma_\gamma}}{E^3} \rho \qquad (4)$$

The level density $\rho$ is to be taken for the upper state of the transition, i.e. the excited levels reached by photon absorption. When the strength function will later be used to describe cascade decay, $\rho$ is meant to describe the region at the upper level in a decay step [5].

In general, radiation absorbed with multipolarity $\lambda > 1$ contributes only weakly, such that one can assume: $\qquad f(E) = f_1(E) + f_2(E) + ... \approx f_1(E) \qquad (5)$

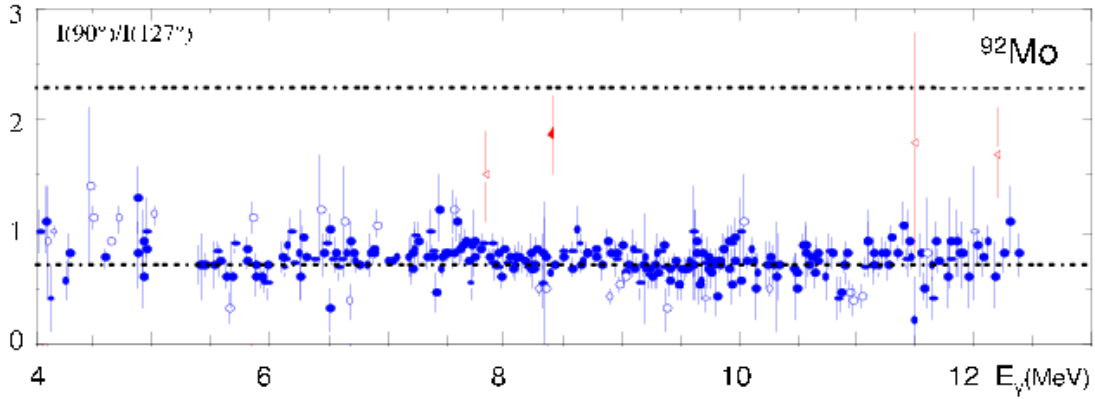

**Fig. 4**: (Color online) Intensity ratios I(90°)/I(127°) for transitions in $^{92}$Mo observed as discrete lines [19, 21]. In the detector geometry used a value of 0.7 corresponds to a dipole transition whereas 2.3 characterizes $\lambda = 2$. Open symbols indicate transitions for which the multipolarity was assigned already previously [16]. The transitions marked in red are considered to be E2.

This is seen for $^{92}$Mo from the intensities of the resolved peaks as observed at the scattering angles 127° and 90° shown as ratios in Fig.4. In the heavier isotopes the distance between photo-excited levels is so small that a clear distinction between spectral lines and quasi-continuum is difficult to be made. This is why the spectra taken for $^{100}$Mo at the two angles are directly presented in Fig. 5. The higher intensity at 127° indicates the dipole character for the peaks and the quasi-continuum in the energy range from 5.5 to 8.5 MeV. At lower energy the spectra are increasingly governed by multi-step gamma transitions causing a loss of directional correlation.

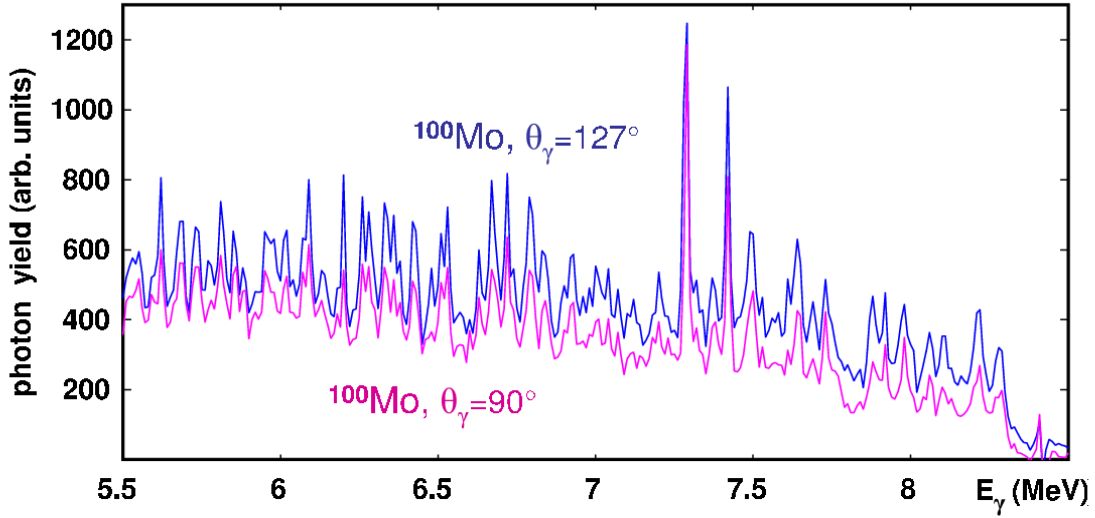

**Fig. 5**: (Color online) Photon yield observed with $^{100}$Mo at 90° (magenta) and 127° (blue). The data are normalized to photon flux and detection efficiency (for the plot both are set to 1 at 7 MeV) and they were corrected for the detector response. In a wide energy range an enhancement at 127° is seen. Because of the spins involved the 7.29 MeV transition in $^{11}$B is expected to show an intensity ratio of 0.93.

We will primarily discuss electric dipole strength, since, as we showed previously [27, 30], the number and strength of M1-transitions is considerably smaller than that of E1- transitions. In a first step of the analysis the response function of the HPGe-BGO combinations is used for a deconvolution of the spectra. Starting at the highest energy all processes not contributing to the full energy peak are subtracted step by step. The procedure was tested by using the HPGe-detectors of this study in a $^{14}$N(p, $\gamma$)-experiment [42] producing spectra with only one high energy line at 10.76 MeV. Its application to data obtained for photon scattering by the low level density nucleus $^{208}$Pb [30] led to results in full agreement with previous experiments [6]. The effect of this deconvolution is seen in Fig.3 for $^{92-96}$Mo and in Fig.3 of reference 19 for $^{100}$Mo.

In a next step the contribution to photon scattering not due to nuclear resonance fluorescence is investigated. Here the background due to multiple Thomson (or Compton) scattering is important and a calculation with the code GEANT3 [35] – modified as discussed above for Fig. 2, with no nuclear resonances included – was used to determine it. As shown in Fig. 3 these processes contribute strongly at low photon energy and as they may exceed 10 % at some energies their contribution had to be subtracted. The energy dependent photon yield $Y(E_\gamma)$ observed for $^{100}$Mo at 90° and 127° is shown in Fig.5; it results from the subtraction of the simulated non-resonant contribution and the correction for the energy dependence of photon flux and detector efficiency and response. After subtracting the peaks coming from ambient background and from the $^{11}$B target the data for all five isotopes are integrated in 200 keV wide bins to get the yields $Y_n$. This binning determines the resolution of the final result, the dipole strength function $f_1(E)$ in dependence of the excitation energy $E$. The next step in the analysis, the correction for feeding and branching, starts from a first guess for $f_1(E)$, from which the (hypothetic) primary populations $I_n$ of the energy bins n – selected to increase in $E_x$ with n – can be calculated by

$$I_n = \sigma_n \varepsilon_n \Phi_n M/A \, N_A = 3(\pi\hbar c)^2 \varepsilon_n \Phi_n M/A \, N_A \, E_n f_1(E_n) \qquad (6)$$

with Avogadro constant $N_A$ and the target mass $M/A$ (in mol). For each bin n $\sigma_n$ is the average cross section, $\varepsilon_n$ the absolute full energy detection efficiency. $\Phi_n$ stands for the primary photon fluence in that bin and it is assumed that the photon intensity is constant over the target volume. From the $I_n$ corrected populations $C_n$ are derived by accounting for feeding and branching – starting at such a high energy where no feeding yield from above contributes:

$$C_n = I_n + \sum_{n>m} \overline{\Gamma_{mn}} \, \rho_n I_m - \sum_{m>n} \overline{\Gamma_{nm}} \, \rho_m I_n = \sum_m A_{nm} I_m \qquad (7)$$

Here $\overline{\Gamma_{mn}}$ is the average width of transitions from bin $n$ to bin $m$ and $\rho_n$ is the level density in bin $n$. Actually this approach would only allow for single step decays but multi-step patterns of order $k$ are easily included without disturbing the linearity of the correction process by applying the matrix $\mathbf{A}$ $k$ times: $\mathbf{C} = \mathbf{A}^k \mathbf{I}$. The ratio $Y_n/C_n$ is a measure to generate the next iteration for $f_1(E_n)$ in an iterative procedure which continues until the changes become insignificant. Starting with a realistic choice for $f_1(E)$ reduces the need for many iterations.

For $\lambda=1$ the average width $\overline{\Gamma_n}$ is proportional to $(\Delta E)^3$ such that one gets by the use of Eq. 4 :

$$\overline{\Gamma_{mn}}\rho_n = f_1(|E_m - E_n|) \ (|E_m - E_n|)^3. \tag{8}$$

Here the Axel-Brink hypothesis [3, 4] was used stating that the strength function $f_1$ only depends on the transition energy $|E_m - E_n|$ and is independent of the energy of the lower level and of the transition direction. Obviously, the average decay from one bin to the other is determined by $E_\gamma$ and $f_1(E_n)$ alone. As in Eq. 6 the primary population is calculated without any other target parameters than $M/A$, $E_n = E_\gamma$ and $f_1(E_n)$ no explicit level density dependence appears; it is completely absorbed in the average quantity $f_1(E)$. Thus our method of obtaining the electric dipole strength does not need any *a priori* information about the level density $\rho$.

The schematic description of the correction for feeding and branching as outlined here does not account for two details which are nevertheless important:
1. A level excited strongly by photon absorption from the ground state has a strong decay branch back to it. When only bins with average quantities are considered this information on the enhancement of elastic scattering [9] is lost.
2. The yields $Y_n$ are observed in an angle limited direction with respect to the photon beam. Thus the spins in the decay cascade and the resulting directional correlations have to be considered as outlined previously [19]. Again this is not done in a bin-wise approach.

If at all, both points disrupt only weakly the statement about the level density independence as made above. But they have influence on the strength function $f_1(E)$ resulting from the data analysis. This is why the bin-wise procedure was not used in the final analysis but instead replaced by a Monte-Carlo description of feeding and branching. In Eq. 7 $\overline{\Gamma_{mn}}\rho_n$ is replaced by $\Gamma_{mn}/E_n$ and the sum now runs over randomly generated individual levels (about $10^3$ in total). The multi-step decay between them observes points 1 and 2; it is completed by a final aggregation of $f_1(E)$ in bins. In this procedure the quasi-continuum resulting from the overshoot of the experimental resolution over the average level distance $D$ is represented by Monte-Carlo generated levels with density $\rho = D^{-1}$. To be independent of the actual choice $10^3$ different realizations of the spectra were used and later the results were averaged. A respective Monte-Carlo-simulation has been the basis for the analysis of an equivalent experiment on $^{88}$Sr [27], where it is described in detail. Also the application of this method to $^{98}$Mo and $^{100}$Mo was already presented [19], where many more details of it are given. It was demonstrated there, that the selection of a specific level density for the Monte-Carlo calculation has an only weak effect on the final results. Results for the five even Mo-isotopes are also shown in ref. 41 and the effect of the correction for multi-step decays on the final strength function is demonstrated in Figure 2 of that paper.

In the decay simulation the levels were randomly generated from a Wigner distribution for the level distances and a Porter-Thomas distribution for their widths. To demonstrate that this approach based on random matrix theory (RMT) is justified, we have shown respective plots for three of the Mo-isotopes before [18, Figs. 3 and 4]. To investigate in detail, if the Porter-Thomas assumption [43] can be made, we selected $^{92}$Mo, which has the lowest level density as compared to the other isotopes. We repeated the RMT based investigation by now making proper account for unobserved levels, which influences the width averaging. Fig. 6 shows the result of that analysis; by using a logarithmic x-axis the presentation is especially sensitive to weak transitions. The 276 discrete transitions observed as peaks in the nrf-spectra above 5 MeV are compared to a $\chi^2$-distribution with one degree of freedom, typical for a Porter-Thomas distribution. The level widths divided by their average are presented for the two photon energy bins 5-9 MeV and 9-12.4 MeV. The averages have been corrected for unobserved lines as well as for the strength hidden in the quasi-continuum: In the lower bin 11 % of the strength is absorbed in the 30 % unobserved transitions and in the higher bin 28 % of the strength and 59 % of the transitions disappeared in the quasi-continuum. No strong singular transitions are observed and an agreement with the Porter-Thomas conjecture can be stated – considering the apparent loss of strength in weak transitions not appearing as discrete lines.

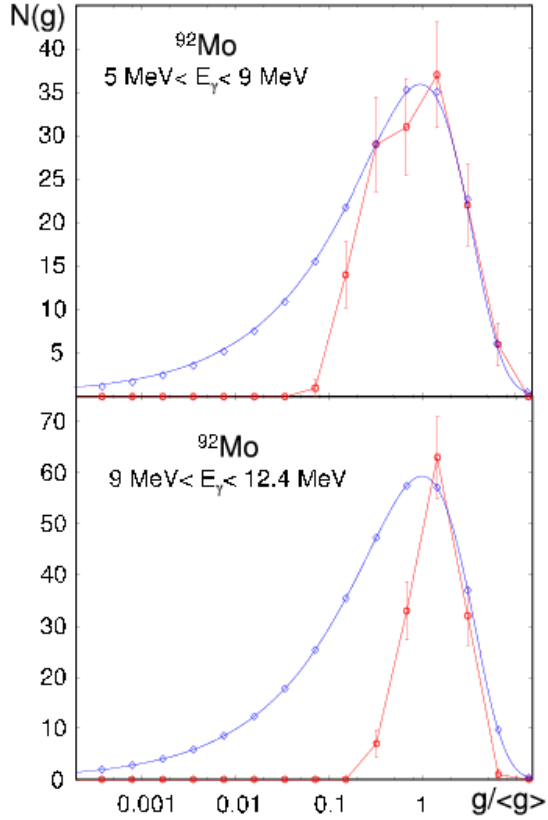

**Fig. 6**: (Color online) For two photon energy ranges the number of discrete transitions in $^{92}$Mo [19] is shown versus the logarithm of their strength $g = \Gamma_0^2/\Gamma$, normalized to the average $\langle g \rangle$. The numbers with their statistical uncertainty are shown in red; they are connected by straight lines to guide the eye. The total number of levels and the full strength in each energy bin are needed to calculate $\langle g \rangle$. The relative strength in the continuum is known [18, 19, 41] and the number of unobserved levels is varied to bring the data at large $g/\langle g \rangle$ into agreement with the smooth curve (blue) depicting the Porter-Thomas distribution. The difference at low $g/\langle g \rangle$, which is more significant at large $E_\gamma$ (bottom), indicates the loss of strength to the quasi-continuum due to weak lines.

Various additional tests were performed within the series of experiments at ELBE to check the reliability of the applied analysis method: In the case of $^{88}$Sr and $^{90}$Zr data were collected not only at 13.2 MeV, but also at two lower energies and consistent results were obtained [27]. For $^{88}$Sr a separate treatment of the statistical decay paths and of individual strong transitions has resulted in a strength function $f_1$ which agrees within the uncertainties. The procedure used has also been successfully applied to $^{90}$Zr [27], and the final results for the absorption cross section for 8-12 MeV compare well to the ones obtained at another electron linac. In contrast to our study tagged photons were used in the same energy range [4] and Fig. 13 below shows the good agreement between the two data sets. In that experiment no correction for feeding is needed and the branching to excited states is explicitly corrected for: The approximation used for the branching correction assumes that the width $\Gamma_c$ is well approximated by a single number typical for the investigated mass region [4, 9]. For the Mo data this alternative analysis method was used also by us [18] in addition to the procedure presented here. Using the same level density and the same $\Gamma_c$ = 0.2 eV proposed for the Zr region [4, 9], the strength functions obtained for $^{92}$Mo, $^{98}$Mo and $^{100}$Mo agree with the results of the present ansatz within the uncertainties. Fig. 7 shows them together with the results for $^{94}$Mo and $^{96}$Mo in comparison to photo-neutron data and to calculations; both will be discussed in subsequent paragraphs.

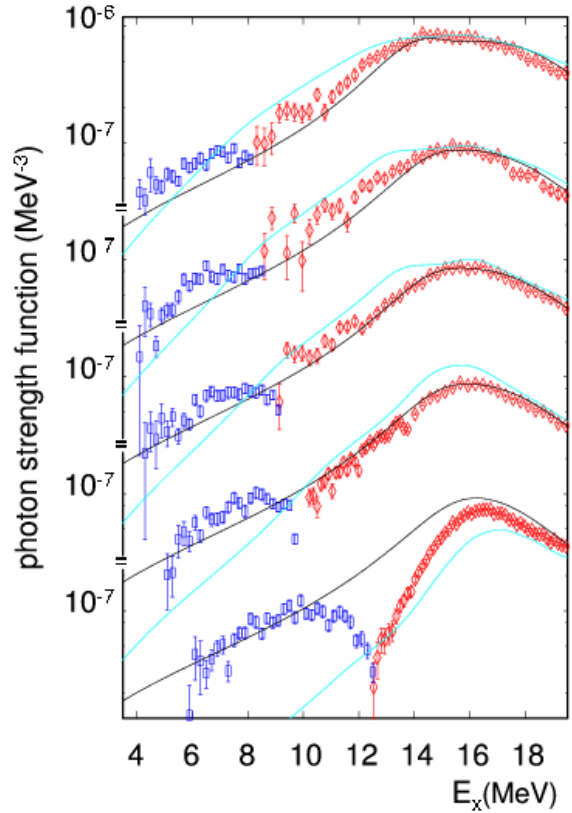

**Fig. 7**: (Color online) Experimental photon strength functions for $^{92,94,96,98,100}$Mo (from bottom to top). The data at low $E_x$ are from the present experiment (□). They are combined with $f_1$ derived from $(\gamma,n)$-data (◊) [20, 44] by making the approximation that photon absorption only leads to neutron emission and by renormalizing them by 0.87 (see text). The black solid lines depict the E1 strength parameterization [22] based on the deformations given in Table I. The thin lines (cyan) show the results of a calculation with the RPA [45] taken from RIPL-2 [23] and discussed under (a) in ch. IV A.

The method applied for the analysis of the scattering data is innovative because of its deconvolution of the spectra to also derive information from their quasi-continuous part, in the procedure of deriving information about the non-resonant processes, as well as in its way to account for multi-step gamma decays by Monte-Carlo-simulation. QRPA calculations for M1 [17] results in corrections < 20 % in the range of 7 MeV and E2 contributions to the yield were shown in Figs. 4 and 5 to be of minor significance. As shown before [19, *cf.* Fig.2], the Monte-Carlo calculations used to simulate the non-nuclear processes are in agreement with the properly normalized data for $^{98}$Mo and $^{100}$Mo in the energy region between the $S_n$ of the two nuclei. A similar conclusion can be drawn from the spectra for $^{94}$Mo and $^{96}$Mo in Fig. 3.

### B. Activation data

Following the simultaneous irradiation of natural Mo and Au with bremsstrahlung of different endpoint energies at the photo-activation site (*cf.* Fig. 1), the gamma-decay induced in the samples was measured in a low background environment. The procedure to obtain absolute activation data was discussed above and published previously [12]. The experimental results for $^{197}$Au($\gamma$,n)$^{196}$Au presented there establish the photo-activation of Au as a reliable standard. After a determination of gamma-transition intensities, and their time dependence, the instantaneous production rates for $^{99}$Mo and $^{91}$Mo were obtained by the comparison to the respective rates for the production of $^{196}$Au. To allow for the comparison of the data obtained for the different endpoint energies, they were normalized to the photon fluence at 7.29 MeV as deduced from a ground state transition in $^{11}$B observed at all electron energies. A direct comparison of our activation results to cross section data obtained previously with monochromatic photons by observing the neutrons [20] would require to determine yield differences from our data. To avoid the resulting large uncertainties we used another approach instead: The numerically available [44] cross section data [20] were folded with normalized bremsstrahlung distributions for the respective electron energy, given by the MCNP-code [36], which is based on published tables [37] – as was discussed earlier. The activation yields for $^{100}$Mo($\gamma$,n)$^{99}$Mo as obtained here agree with the old ($\gamma$,n)-data [20] from Saclay only after these are multiplied by 0.89 ± 0.04, a factor determined from a $\chi^2$-fit by using the data displayed in Fig. 8. To demonstrate the constancy of this factor, it is shown in the bottom part of Fig. 8 in its dependence on the endpoint energy. Due to the reference to $^{11}$B at 7.29 MeV [34] the absolute normalization accuracy is 7 %. A factor of 0.85 ± 0.03 had previously been shown [11] to be necessary for an optimum agreement between

Saclay data for Zr($\gamma$,n) and several neighboring nuclei by a high accuracy experiment performed by an international collaboration at the Lawrence Livermore Laboratory. Consequently we applied the average, *i.e.* a factor of 0.87± 0.05 to the Saclay data [20] for the five Mo isotopes.

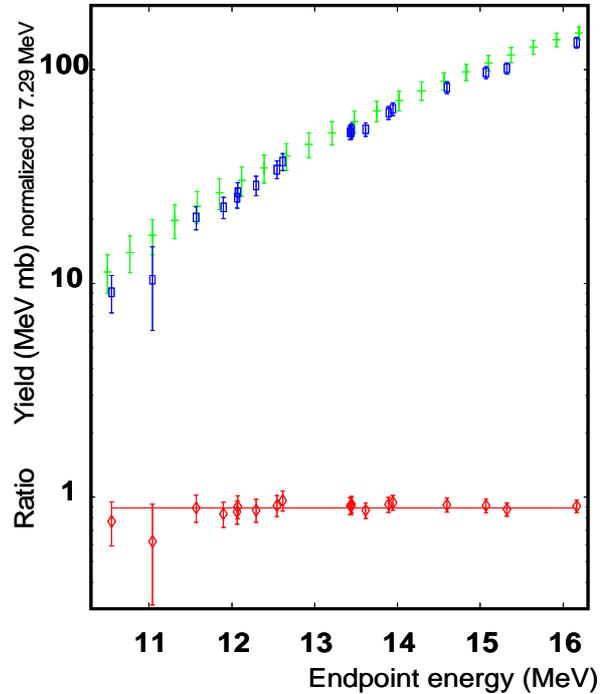

**Fig. 8:** (Color online) Activation yields observed for $^{100}$Mo($\gamma$,n)$^{99}$Mo (□, blue) in dependence on the bremsstrahlung endpoint energy; they were normalized to the flux observed at 7.29 MeV. Previous cross section data (+, green ) [20, 44] and, in the bottom of the Figure, yield ratios from this experiment compared to them are also shown (◊, red) with their uncertainties; a line at 0.89 depicts the average ratio. For the comparison the previous data were integrated and interpolated accordingly.

The corresponding data for the $^{92}$Mo component in the irradiated samples are especially interesting for the present strength function study, as for that nucleus the neutron separation threshold is more than 5 MeV higher than the proton threshold. In the interesting energy range the photon absorption is thus connected to four exit channels: ($\gamma$,n), ($\gamma$,p), ($\gamma,\alpha$) and ($\gamma,\gamma'$). The first three are accessible by our activation technique, when taking account of the decay of the short lived $^{91}$Mo to $^{91}$Nb. This has to be separated properly from the yield of $^{91}$Nb directly produced via $^{92}$Mo($\gamma$,p). With our pneumatic delivery system such a separation was possible and it allowed to gain results for the population of the 65 s spin isomer in $^{91}$Mo. These data (*cf.* Fig. 10) will be discussed together with statistical model calculations of Hauser-Feshbach type in the following. Recent publications [2, 25, 31, 45] address the relevance of results like the ones given here to cosmic nucleosynthesis calculations for the region of Mo.

# IV. DIPOLE STRENGTH AND THE GDR
## A. Microscopic description

The isovector giant dipole resonance (GDR) was the first established prominent collective nuclear excitation, and therefore has attracted wide interest. There have been numerous theoretical descriptions, many of which can be extended to the low energy tail which is of main interest here. Starting from a microscopic approach (as alternative to earlier macroscopic descriptions) Brown and Bolsterli have worked out in their seminal paper [46] of 1959 that already in a harmonic oscillator basis, and then also in the nuclear shell model, a strong dipole excitation is expected at 10 -15 MeV. By adding various residual nucleon-nucleon-interactions to the mean field a large variety of calculation schemes has been developed – mainly for spherical nuclei. Practically all of them are limited to first order, *i.e.* they produce one-particle one-hole states only and treat two-particle two-hole configurations (if at all) by including collective quadrupole degrees of freedom. The resulting mean level distance exceeds the experimental value by more than a factor of 10 and thus an arbitrary smearing parameter has to be introduced to conceal that deficit. A smearing width of ~ 2 MeV is often used to obtain a smooth energy dependence of the cross section. Three calculations dedicated explicitly to the dipole strength function in the Mo-isotopes have been presented:

a) By a random phase approximation (RPA) with a density dependent interaction, a very large number of heavy nuclei have been studied [45], and respective predictions have been presented together with the RIPL-2-compilation [23] of photon strength function data. There it is stated : *"The E1-strength functions are determined within the QRPA model based on the SLy4 Skyrme force. The ground state is consistently calculated within the Hartree-Fock+BCS model based on the same SLy4 force. . . . The QRPA calculations are performed in the spherical approximation. A folding procedure is applied to the QRPA strength distribution to take the damping of the collective motion into account. In the case of deformed nuclei, a phenomenological splitting of the QRPA resonance strength is performed in the folding procedure"*. Here the deformation corresponds to the ground state mass minimum in QRPA and the inclusion of deformation effects appears to be somewhat arbitrary. Results from this calculation are displayed in Fig. 7; they differ significantly from the combined experimental data.

b) A 2$^{nd}$ approach [47] uses a potential energy surface calculation to obtain the deformation parameters (*cf.* Table I) and a deformed basis obtained from the Nilsson model to carry out a

RPA. As residual nuclear interaction a density independent dipole-dipole force is used. To get the resonance energies right, an additional arbitrary adjustment is needed. In a first publication [47] calculations for Kr, Mo and Nd isotopes are presented as cross section integrals over 10 MeV. These numbers agree with the experimental equivalents within up to 35 %; for $^{148}$Nd – but not for $^{150}$Nd – this comparison favors a deformed mean field over a spherical one. For the Mo isotopes the integrals in the slope region agree with the ELBE-data within their experimental uncertainties; for the integrals including the GDR peaks the experimental values are exceeded by ~ 30% in $^{98}$Mo and $^{100}$Mo. Calculated strength distributions in the GDR region are given [ref. 47, Figs. 3 and 7] for Nd and Mo, respectively. They show two peaks: the higher energy one is considerably narrower than the experimental GDR peak and the smaller one at ~ 4.5 MeV below the main peak has < 30 % of its strength. As it considerably exceeds the data in this energy range the cross section near $S_n$ is likely to be due to this side-peak the tail of which is determined by the smearing width. For the even Mo-isotopes more results have been published recently [41]; here the tail region and not the GDR peak are discussed.

c) A density functional RPA formalism, which had already been used for several regions of the nuclide chart including permanently deformed nuclei [48], has recently been applied to the Mo-isotopes [49]. By using interaction parameters of Skyrme type, some agreement with data is obtained, but only axial deformations were studied. Apparently the results differ from (b) and it is stated: *"The low-energy E1 strength is shown to be mainly determined by the GDR tail"*. This statement is in agreement with the findings of the present work, as will become obvious below. As the predicted position of the GDR deviates from the maximum in the ($\gamma$,n) data [49] a direct comparison to our ($\gamma,\gamma$) data appears to be premature. It is not obvious if a final selection between different choices for the interaction will lead to a satisfactory global description of many nuclei.

Very recently, a residual interaction, derived using the newly developed 'Unitary Correlation Operator Method' has been used [50] in second order RPA calculations for the neighbouring $^{90}$Zr nuclide. The explicit inclusion of 2-particle 2-hole configurations results in a rather weak low energy side-peak with consequently less impact on the slope at $S_n$ as compared to other calculations. An interesting feature of this work is the employed response function formalism which allows to introduce a smearing width in an early step of the calculation. An interesting new development in the field of RPA calculations is the use of mean field parameters derived from covariant density functionals [51]. These calculations describe many aspects of dipole strength functions in spherical nuclei very well, but their expansion to nuclei without spherical symmetry has not been published yet.

### B. Macroscopic picture

From the time of its discovery, the large strength of the GDR has triggered attempts to describe it as a collective mode, and the out-of-phase oscillation of protons and neutrons around the center of mass has become the generally accepted macroscopic description. A single peak observed at 10-15 MeV in nuclei with at least one magic nucleon number indicates one oscillatory mode as expected for a spherical body. For a triaxial shape three components of equal strength are to be seen, and in an axial nucleus 2 of the 3 components coincide. Again this is confirmed by the observation that in nuclei known to be axially deformed two peaks appear, which are separated by up to 4 MeV. A model derived on this basis by Goldhaber and Teller [52] works well for lighter nuclei, whereas the one of Steinwedel and Jensen [53] gives a better description for very heavy nuclei. On the basis of the droplet model [54-55] unified expressions were developed to hold for both mass ranges. Thus a prediction for the A and Z dependence of the centroid energy of the GDR can be obtained by combining one new parameter – an effective nucleon mass – to the nuclear symmetry energy and surface stiffness already determined [55] by a fit to ground state masses. Concerning the energy dependence of the apparent GDR width, a rather simple formula was derived [56] on the basis of hydrodynamical considerations, which describe it as a spreading width. Originally it was proven to hold within one nucleus, but it was empirically shown [22] that this expression can be generalized to be valid for all heavy nuclei (A>80) with only one free fit parameter. Likewise it was found recently [22] from inspecting the isovector GDR in many heavy nuclei, that its integrated strength exhausts the TRK sum rule without any addition to its main component, which is proportional to (Z·N)/A. This rule was formulated by Kuhn as well as by Reiche and Thomas [57] for electric dipole absorption by any charged microscopic object. Later is was shown that the summed photon absorption cross section in nuclei can be quantified already by only using a very general concept based on causality and unitarity [58]. Applying the abovementioned considerations a Lorentzian parameterization for the dipole strength was derived [22] and for many nuclei with A>80 it was shown to well describe the peak area of the GDR up to ~ 3 MeV above, when the $T_>$-component starts to contribute to the cross section. A possible extension of this ansatz to lower energies will play a major role in the subsequent discussion. For the lower energy regime well below the GDR it is known that E1 transitions are strongly over-predicted by the single particle Weisskopf unit – in contrast to other multipolarities.

Apparently the electric dipole strength is concentrated in the GDR, and already many years ago the assumption was introduced [4, 5] that the low energy tail of the GDR determines the strength of absorption and emission of photons in this energy range. But while nuclear fluorescence data were well described by such an extrapolation [4, 5, 9], the experimentally determined gamma decay strength following neutron capture seemed to indicate a steeper energy dependence as expected from a Lorentzian fitted to the GDR peak. Inferring from the concept of an excitation energy dependent width [56, 59], that the GDR width decreases with the photon energy [2, 23-24] caused disagreement between the nrf and the capture communities, and induced a longstanding controversy [23, 24, 60]. By a careful inspection of the ($\gamma$,n)-data used for the determination of the GDR parameters, and by taking a consequent account of the nuclear deformation and triaxiality a solution to this puzzle was found: A parameterization of the electric dipole strength originally developed mainly to describe nrf-data as well as the GDR itself could be shown [22] to be in good agreement to many capture data, namely those obtained by average resonance capture (ARC) of neutrons. In contrast to other prescriptions [23-24, 60], it is based on a spreading width only depending on the resonance energies [59] and not on the photon energy directly. The potentially higher strength at low energy is reduced due to the use of a smaller spreading width. This, in turn, is possible, as a considerable portion of the apparent width is attributed to the nucleus' deformation. An especially sensitive test of this ansatz is possible for nuclei, for which reliable data at low and at high energy exist. For a considerable number of heavy nuclei sufficient information was available and the electric dipole strength was explained well [22] by this new parameterization. As already indicated in Fig. 7 a comparison to the low energy data obtained in this study in combination with the Mo($\gamma$,n)-data [20] for the GDR region is of special interest. Special insight in the photon energy dependence of the width is expected from the stable even Mo nuclei with their wide variation in deformation. These data do not cover the regime below 4 MeV and thus do not allow any conclusion on the prediction made of an extra strength at such low energy, based [60] *e.g.* on the theory of Landau liquids. One other relation, the photon energy dependence of the width, is often discussed [23-24] in parallel, but we argue that a distinction should be made between these two predictions. Their origin is not directly related and they mainly effect different excitation energy regions.

For the deformation parameters, different sources are available for the Mo-isotopes (*cf.* Table I): Experimentally, the low energy spectrum of $^{92}$Mo does not show a sign of strong static deformation, and for $^{98}$Mo a detailed Coulomb excitation study [26] has delivered rms-values for the deformation parameters $\beta = 0.18$ and $\gamma = 33°$ indicating a dynamically non-axial shape. The derivation of $\beta$-values from B(E2)-values to the first excited $2^+$ state [61] results in a rather good approximation (actually a lower limit) [26] for the rms deformation $\beta$. We use these values in combination with a systematic data analysis [62], which also published results for the triaxiality $\gamma$ in correlation to the rms deformation $\beta$. A surprisingly clear dependence between the two observables is shown, and the proportionality constant relating them in the even Mo isotopes is similar for other heavy nuclei. We applied this information to derive the values shown in the last two columns of Table I, which we used in the comparison of strength functions from ref. 22 to our data. The sensitivity to possible uncertainties in these values will be demonstrated at the end of ch. V B. Theoretical deformation parameters included in the Table for comparison come from the finite range droplet model (FRDM) tabulated for more than 900 nuclides, recently with inclusion of triaxiality [55], and calculations of Nilsson-Strutinsky type (NS) [17, 47]. The column labelled ISS shows the weighted averages of deformation parameters assuming a slowly (as compared to the dipole oscillation) changing deformation. They were determined [63] from properties of low lying levels using the interacting boson model (IBA). In the following this information is used to demonstrate the impact of a time dependent coupling of the low energy quadrupole vibrations to the giant dipole mode.

**TABLE I**: Deformation parameters for the Mo-isotopes taken from: B(E2) values assuming axial symmetry [61], a Coulomb excitation experiment [26], the FRDM [55], a Nilsson-Strutinsky (NS) calculation [17] and the IBA [63]. The last two columns show the values selected here; they are based on [26], [61] and [62] (*cf*. Fig. 12).

| nucleus | | axial | Coulex | | FRDM | | NS | | IBA/ISS | | used here | |
|---|---|---|---|---|---|---|---|---|---|---|---|---|
| N | A | $\beta$ | $\beta$ | $\gamma$ | $\beta$ | $\gamma$ | $\beta$ | $\gamma$ | $\beta$ | $\gamma$ | $\beta$ | $\gamma$ |
| 50 | 92 | 0.10 | | | 0.03 | 0 | 0 | 0 | 0.10 | 24.9° | 0.11 | 34° |
| 52 | 94 | 0.15 | | | 0.05 | 0 | 0.02 | 60° | 0.14 | 19.7° | 0.15 | 31° |
| 54 | 96 | 0.17 | | | 0.19 | 32.5° | 0.11 | 60° | 0.16 | 26.2° | 0.17 | 29° |
| 56 | 98 | 0.17 | 0.18 | 32° | 0.22 | 27.5° | 0.19 | 37° | 0.20 | 22.7° | 0.18 | 25° |
| 58 | 100 | 0.23 | | | 0.24 | 25.0° | 0.23 | 32° | 0.25 | 22.4° | 0.23 | 22° |

From the deformations $\beta$ and $\gamma$ the GDR resonance energies are calculated to be inversely proportional to the respective diameter of the triaxial body using the Hill-Wheeler formula [22]:

$$E_k = E_0 \, R_0 / R_k = E_0 / \exp[\sqrt{5/4\pi} \, \beta \cos (\gamma - \tfrac{2}{3} k \pi)], \tag{9}$$

with $E_0$ calculated from the symmetry energy in the finite range droplet model (FRDM) [55], which is fitted to the ground state masses of many nuclei. Now equation (1) can be used here in a different sense than before, when it described the integration over the narrow (~eV) resonances, $\int_\Delta \sigma dE = \Delta <\sigma>$ to yield an average cross section, which is proportional to $E \cdot f_1$ when forming the sum in (1). For the description of the GDR, the sum is now taken over (up to three) Lorentzians corresponding to the different radii $R_k$ for the (eventually triaxially) deformed nucleus (*cf.* Eq. 4 in [22] and Eq. 9 here). Here the widths (now in the MeV range) no longer account for the level's decay into a vacuum but rather for its spreading into the underlying narrow levels of multi-particle multi-hole structures. A description by Lorentzians has been proven to be valid also for that case [59], and the resonance parameters for the (up to three) components of the GDR are given by [ref. 22, Eq. 3 and 5]:

$$\Gamma_k(E_k) = 0.05 \, E_k^\delta, \, \delta = 1.6 \qquad (10)$$

The resonance energies and widths, derived from the shape parameters in the last columns of Table I, are given in Table II. The dependence [59] of the widths on the resonance energies $E_k$ and the exponent appearing in (10) were derived from hydrodynamical considerations [56], and the proportionality constant stems from a global fit [22] to many nuclei with A>80.

## V. RESULTS

### A. Strength functions in the Mo isotopes and their parameterization

The even Mo isotopes selected here constitute an interesting chain of nuclei as they vary in deformation as well as in triaxiality [17, 26, 55]. The ($\gamma$,n) cross sections are available [20, 44] for all of them, and our activation studies have confirmed the earlier finding [11] that their absolute value has to be reduced by approximately 13 %. In the following all ($\gamma$,n) data are corrected respectively. Especially for $^{92}$Mo with its exceptionally large $S_n$= 12.7 MeV an additional correction to the data is important: Near 13 MeV the – albeit small – target admixtures of other Mo isotopes had to be subtracted as they contribute considerably to the yield which, for $^{92}$Mo, is very small due to the close threshold. This was accomplished on the basis of the procedure, described below, used to calculate cross sections. As described above, the deformation parameters entering the parameterization [22] of the dipole strength function are taken from published information [26, 61, 62]. The parameters resulting for the different isotopes are listed in Table II, together with the respective values as documented in RIPL-2 [10, 23] for a description by a single Lorentzian. Whereas the deformation induced split is quite marginal for $^{92}$Mo, it increases considerably with increasing neutron number.

**TABLE II**: GDR-Parameters used for the description of the E1 strength in the Mo-isotopes studied. Given are the energies and widths (both in MeV) of the up to three Lorentzians and their strength relative to the TRK-sum (in%). Following ref. 22 the GDR-components used here (columns 5-7) have equal strength which adds up to the TRK-sum. In comparison the values proposed in RIPL-2 [23] are shown. They are taken from Dietrich and Berman [10], and are not in good agreement with the TRK-sum rule [10, Eq. 3].

|     | RIPL-2 [23] | | | ref. 22 | | |
| --- | --- | --- | --- | --- | --- | --- |
| A   | $E_R$ | $\Gamma_R$ | % | $E_R$ | $\Gamma_R$ | % |
| 92  | 16.82 | 4.14 | 77  | 15.72 | 4.11 | 33.3 |
|     |       |      |     | 16.54 | 4.45 | 33.3 |
|     |       |      |     | 17.65 | 4.94 | 33.3 |
| 94  | 16.36 | 5.50 | 115 | 15.27 | 3.92 | 33.3 |
|     |       |      |     | 16.53 | 4.45 | 33.3 |
|     |       |      |     | 17.99 | 5.09 | 33.3 |
| 96  | 16.20 | 6.01 | 123 | 15.01 | 3.81 | 33.3 |
|     |       |      |     | 16.53 | 4.45 | 33.3 |
|     |       |      |     | 18.11 | 5.15 | 33.3 |
| 98  | 15.80 | 5.94 | 122 | 14.87 | 3.75 | 33.3 |
|     |       |      |     | 16.61 | 4.48 | 33.3 |
|     |       |      |     | 18.03 | 5.11 | 33.3 |
| 100 | 15.74 | 7.81 | 144 | 14.43 | 3.54 | 33.3 |
|     |       |      |     | 16.74 | 4.54 | 33.3 |
|     |       |      |     | 18.40 | 5.28 | 33.3 |

The strength functions $f_1$ as calculated using these resonance parameters are depicted in Fig. 7. The experimental low energy strength functions, as derived from our scattering data following the procedure described above (eqs. 1,6,7,8), are also shown. These experimental results were already published earlier [19, 41] and are shown again here. To get an impression of the excitation energy dependence of the dipole strength in a wider range, the ($\gamma$,n)-data [20] for $E_x >$ $S_n$ are included in the Figure. The cross sections were transformed into $f_1$ after adjusting them by 0.87 (*cf.* Fig. 8 and the corresponding text) and by using Eq. 4. Only in the case of $^{92}$Mo, which has a very low proton emission threshold $S_p$ and thus a strong ($\gamma$,p) channel, the assumption that this channel may be neglected in (2) is not a good approximation. This can be clearly seen in Fig. 7 and a proper treatment of this point using a statistical reaction model will be discussed below. In the other isotopes, the scattering data from close below $S_n$, match reasonably well the strength obtained from the ($\gamma$,n) data [20] directly above $S_n$. The ground state deformations shift the lowest of the three GDR resonance energies away from the centroid by up to 2 MeV and subsequently they also modify their widths by up to 22 %. As seen in Fig. 7 this results in a shape different from, albeit quite similar to, a Lorentz curve, and explains why past attempts [10] to use a fit with a single Lorentzian are apt to lead to unreasonably large width values. These were eventually interpreted incorrectly as a spreading width and this resulted in a false extrapolation to the low energy tail. In our parameterization [22] the rather good description of the data below the neutron threshold is the result of a shift of one component of the GDR to lower excitation energy, and thus an increase of the strength there. No explicit photon energy dependence of the resonance width is needed and all parameters entering (except the deformations) are resulting from a procedure considering globally all heavy nuclei. It may be stressed here again, that we have not performed any fit to the Mo-data to obtain the quite appealing agreement with the experiments. The deformation independent global parameters are valid for all A>80; they are:

1. an effective nucleon mass of 874 MeV combined with

     the symmetry-energy $J = 32.7$ MeV and the effective surface-stiffness constant

     $Q = 29.2$ MeV, which are already fixed by a fit to ground state masses [55];

2. the proportionality factor in Eq. 10 which is used together with the exponent 1.6 already fixed

     by a hydrodynamical calculation [56].

As outlined before [22], we use only these numbers to obtain all relevant GDR energy parameters. We combine them to the strength information derived from the dipole sum rule [57, 58], from which we get the three resonance integrals $I_1=I_2=I_3$ used in eq. 11 below. For the higher energies above the thresholds the relative transmission coefficients for neutrons and protons may have to be determined additionally, as discussed in the next section.

The procedure used here to determine the GDR energies and widths for the calculation of the photon strength functions is valid for a static triaxial deformation. A fully adiabatic coupling of the fast GDR motion to the low-frequency collective excitations, responsible for the dynamics of the ground state deformation, may cause differences. Up to this point these were assumed to be negligibly small, and the root mean squares (rms) of the deformation parameters were treated as static values. To implement a less simplified coupling of the GDR to the quadrupole dynamics of the nuclear body, it is interesting to study the effect of the variance of the deformation parameters as, *e.g.* , observed recently [26] in Coulomb excitation. It has been proposed recently [63], how a description of such dynamics can be derived within the interacting boson model (IBA). This procedure allows to introduce an instantaneous shape sampling (ISS), and it was worked out originally [63] for a coupling to a GDR generated within a RPA. Obviously the ISS can also be combined to a Lorentzian description as proposed here: This requires to calculate the three GDR components at each sampling point and to form the time averaged cross section. To quantify the change introduced by the ISS this average has to be compared to the cross section obtained in a 'single shot' with the average deformations from the ISS (*cf.* Table I). Such a comparison was performed using the time dependent shape parameters as taken from spectroscopic information for the five Mo isotopes [63, *cf.* Fig.1] and the corresponding result is depicted in Fig 7. ISS causes some change near the GDR peak for the triaxial $^{100}$Mo but its impact in the region of our nrf-data is of minor importance.

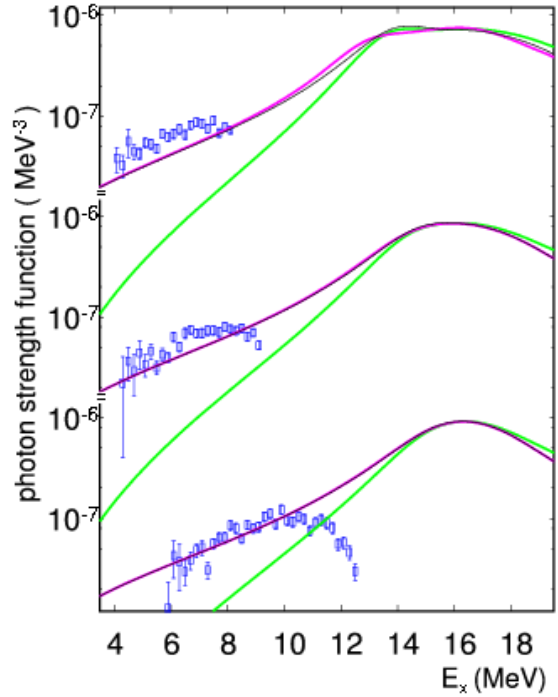

**Fig. 9**: (Color online) Calculated dipole strength functions for $^{92,96,100}$Mo (from bottom to top). Thick pink line: Weighted sum of Lorentzians with deformation parameters as derived [63] by instantaneous shape sampling (ISS) on the basis of the IBA. Black thin line: The deformation parameters were first averaged and then used in the Lorentzian description of the dipole strength; see Table I. The lower thick green line shows the Lorentzian parameterization with an explicitly introduced quadratic dependence on $E_\gamma$; for the $^{94}$Mo and $^{98}$Mo respective comparisons are graphically presented in [22]. An agreement with $f_1$ as determined from the present nrf-data (blue squares with error bars) is only found for the curves calculated without photon energy dependence.

Figure 9 also displays the experimental motivation for not introducing a photon energy dependence of the GDR widths: Curves calculated with $\Gamma \propto E^2$ clearly underpredict the data. To come to a quantitative statement for a possible dependence of the GDR widths on photon energy we made an equivalent ansatz as used in Eq.10 for the resonance energy dependence. We find that any exponent > 0.3 can be clearly excluded by our low energy data. Noting that the spreading width is dominating the GDR width, we conclude that our analysis of the combined data from above and below threshold indicates: The spreading width depends on the resonance energy only, and not on the energy of the emitted photon. This finding contradicts previous practice [2, 23, 24]: The effect of the large GDR width (*cf.* column 3 of Table II) at low energies was compensated by introducing a photon energy dependence of that width.

### B. Comparison to Hauser-Feshbach calculations

In Figure 10 the nuclear photo-effect channels which contribute strongly to the photo-absorption of $^{92}$Mo are shown. The data are compared to calculated photoactivation yield curves, that are produced from Hauser-Feshbach statistical model calculations performed with the code TALYS [64]. Transmission coefficients obtained from the optical model for nucleon-nucleus interactions and the photon strength $f_1$ resulting from the parameterization [22] are used as input for these calculations. The strength $f_1$ – and the average photo-absorption cross section $<\sigma_\gamma(E)>$ – are related [5] to the GDR parameters via :

$$<\sigma_\gamma(E)> \equiv 3(\pi\hbar c)^2 E f_1(E) = \sum_{R=1}^{3} \frac{2 I_R}{\pi} \frac{E^2 \Gamma_R}{(E_R^2 - E^2)^2 + E^2 \Gamma_R^2} \; , \quad (11)$$

which is equivalent to eq. 1. The energies $E_R$ and widths $\Gamma_R$ are given by Eq. 9 and 10 and listed in Table II. As shown before [12, 22], our parameterization allows to describe the E1 strength for nuclei with A>80 in accord to the TRK sum rule [57]. Thus the three resonance integrals $I_R$ in (1) add up to 6.0·(Z·N)/A; we set them to be equal and get $I_R$ = 2.0·(Z·N)/A for R=1,2,3 (cf. Table II). The green and blue curves correspond to optical model parameter sets [65-66] extracted in two different ways from particle scattering data. Using an optical model, the dependence on the relative level densities in the final nuclei $^{91}$Nb, $^{91}$Mo and $^{88}$Zr is accounted for. From inserting the different options [65-66] installed in the TALYS code [64, see also 23, RIPL-2] this dependence was found to be weak, as demonstrated in Fig. 10. However, an intermediate structure detected in a electro-proton experiment on $^{92}$Mo [67] indicates a non-continuous level density or transmission coefficients to be present in $^{91}$Nb. Such a structure may induce a reduced ($\gamma$,p) cross section. Because of flux conservation, a 25 % reduction of this cross section in the $\gamma$-energy range between 11 and 13 MeV may easily cause a much stronger increase in the weaker ($\gamma$,n) and ($\gamma,\gamma$) channels, such that the apparent differences to the experimental data in Figs. 10 and 11 disappear.

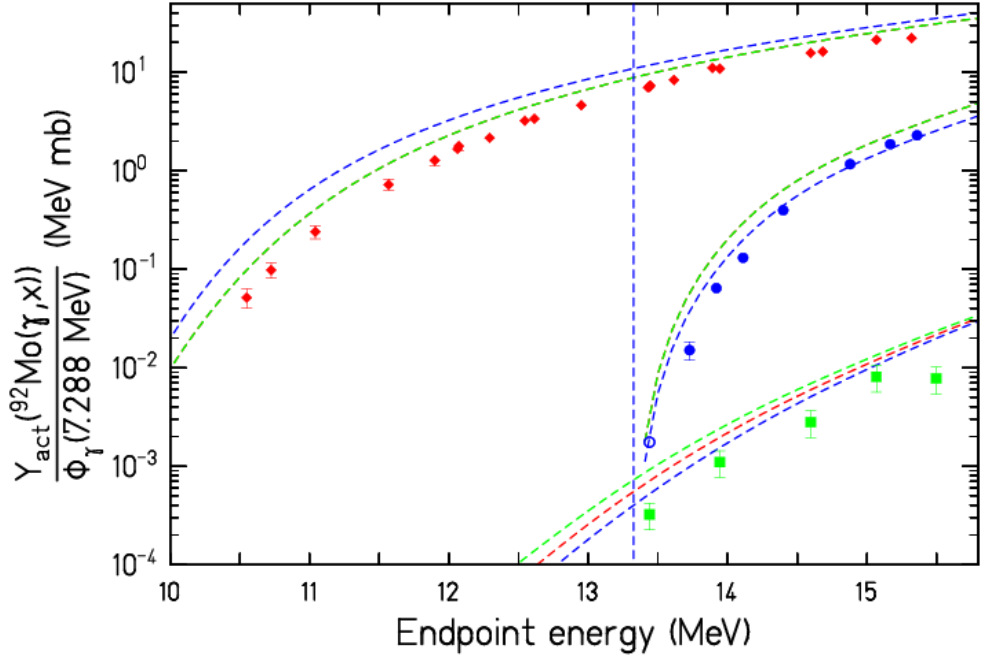

**Fig. 10**: (Color online) Experimental activation yields for the bremsstrahlung induced photo-disintegration of $^{92}$Mo into $^{91m}$Nb, $^{91m}$Mo and $^{88}$Zr (from top to bottom). These channels correspond to the photo-effect resulting in a free proton, neutron or $\alpha$-particle. The blue open circle indicates the detection limit of the used set up for $^{91m}$Mo. The lines depict results of Hauser-Feshbach calculations performed with the code TALYS [64] and the present strength function parameterization as input combined to different optical potentials [65-66] to show the weak sensitivity on their choice. The dashed vertical line indicates the ($\gamma$,n)-threshold for the isomer $^{91m}$Mo.

A more detailed discussion of the influence of optical model parameters and transmission coefficients on competing cross sections has been presented earlier [31]. It is also described there, how strong Hauser-Feshbach calculations depend on the photon strength function: Its size influences essentially linearly the calculated cross sections and yields of photo-nuclear processes. An approach proposed previously [code NON-SMOKER, ref. 2] to use $\beta$-deformation values resulting from the droplet model [55] for the derivation of a two-component Lorentzian do not describe the GDR's in the Mo isotopes [31] as well as the three-component curves [22] described here. In addition, the method used for the calculation of the GDR widths and their dependence on the photon energy is not in accordance to the procedure [22] used here for the prediction of the photon absorption cross section in the range between 4 MeV and $S_n$. Thus an agreement with the photon scattering data presented here is unlikely; respective results from NON-SMOKER [2] are not available at present.

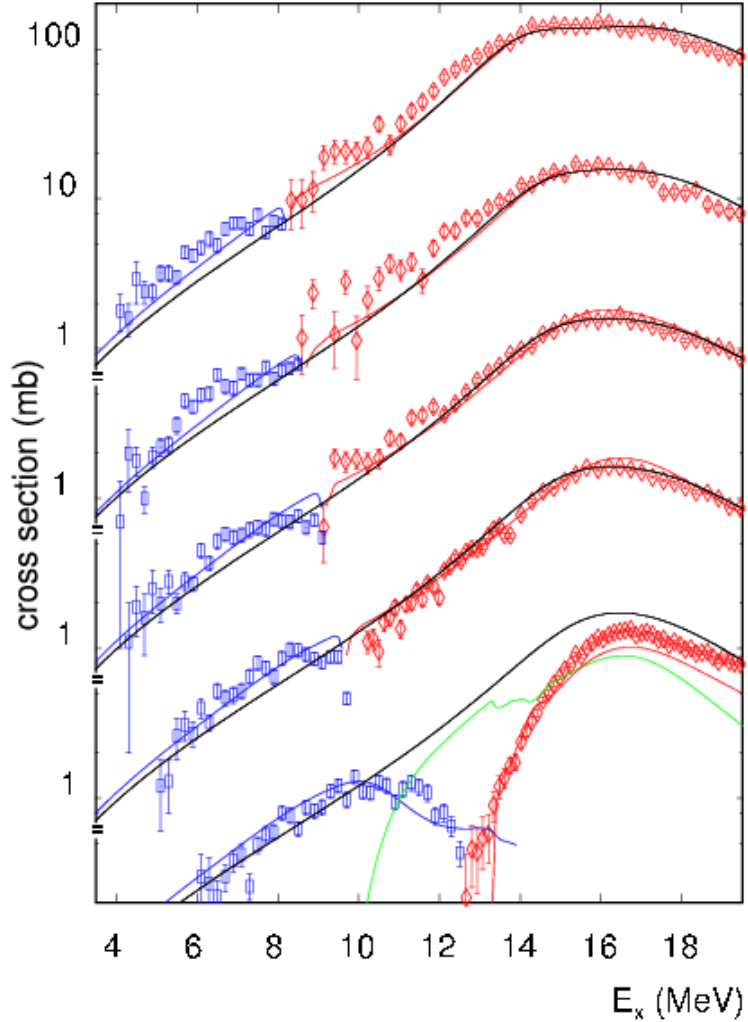

**Fig. 11**: (Color online) Experimental cross sections for photon induced processes in $^{92,94,96,98,100}$Mo (from bottom). The data at low $E_x$ from the present scattering experiment (blue □) are shown together with $(\gamma,n)$-data (red ◊) [20, 44] rescaled as described in the text. The thin lines depict the results of Hauser-Feshbach calculations performed with the code TALYS [64] (blue: $(\gamma,\gamma)$; red: $(\gamma,n)$; green: $(\gamma,p)$; only shown as long as their contribution exceeds 10%). These calculations use the parameterization [22] for the absorption cross section $\sigma_\gamma(E1)$, represented by the thick solid line, which is based on the deformations given in Table I. A cross section overshooting this line may indicate contributions from M1 or E2 as included in the TALYS code [64].

Figures 11 and 13 show the cross sections for photon scattering (nrf) as well as the ones for the photon induced processes of importance for the present discussion. The experimental data agree reasonably well to the cross sections for the different exit channels as calculated with the code TALYS [64] by exchanging the 'standard' electric dipole strength functions of that code by the one taken from [22] and outlined above. In the spirit of the statistical reaction model an average over all possible paths to the same final nucleus have to be performed as well as a summation over all the narrow (~ 1 eV) resonances $R$; this is implicitly done by the code TALYS. It uses transmission coefficients calculated in an optical model and takes into account the level densities in the participating nuclei. As long as only one process contributes to the absorption at a given photon energy, the data may directly be compared to the parameterization for the dipole strength [22], as used here. As shown in Fig. 11 this is not the case for $^{92}$Mo for most of the energy range, and the apparent mismatch between ($\gamma$,n)-data and TALYS near the GDR maximum of $^{92}$Mo may well originate from a too high Hauser-Feshbach prediction for ($\gamma$,p) as already apparent in Fig. 10. In the other isotopes the situation is much less complex and no such uncertainties enter in the calculations. The ($\gamma$,n) data for the low energy slope of the GDR show very large irregularities in the case of $^{98}$Mo; the fluctuations observed in another experiment [32] on the same nucleus are much weaker. Thus a verification of the irregularity in $^{96}$Mo($\gamma$,n) just above $S_n$ would interesting, but no alternate data were found for $^{96}$Mo [44] and our activation method cannot be applied. Apparently, the M1 strength included in the TALYS code improves the agreement to the data, if it has a significant effect at all. It has been predicted that spin-flip M1 strength should be comparatively strong at ~ 8 MeV [23], and orbital M1 excitations have been reported near 3 MeV for many heavy nuclei [15].

To demonstrate the dependence of the calculated cross sections on the deformation parameters, it is advantageous to select cases with only one strong exit channel, which then is a direct measure of the photon absorption. Fig. 12 shows the GDR peak region for the Mo-isotopes 94 and 96, where the absorption cross section is nearly completely exhausted by neutron emission. For $^{94}$Mo the ($\gamma$,n)-cross section is well described by our choice of the deformation parameters, as listed in the last columns of Table I and described in the associated discussion. The calculation based on the values derived by the FRDM [55] shows a reduced strength in the GDR wings and an excess in the peak, an effect of the resonance integral being predefined by the TRK sum rule. The $^{94}$Mo($\gamma$,n)-data near the GDR maximum clearly allow to exclude a fully prolate or oblate shape and show preference for the two triaxial alternatives; the data do not allow to distinguish between these two. Concerning the region near and below the neutron separation energy $S_n$, i.e. below 11 MeV, the value of $\gamma$, defining the triaxiality, has a negligible direct effect on the cross section. The triaxiality is of indirect, but nevertheless great, importance: For a long time its disregard has had the consequence, that only in the case of a clearly double humped GDR the ground state deformation was identified as the cause for its broadening. In nuclei like $^{94-100}$Mo the broadening was misinterpreted as being due to an increased spreading (see Table II, column 3).

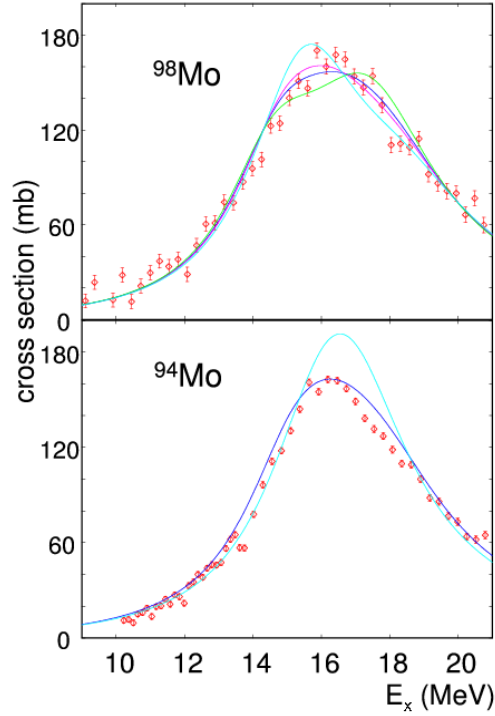

**Fig. 12**: (Color online) Experimental cross sections for ($\gamma$,n) in comparison to the parameterization with different choices for the deformation values. Bottom: $^{94}$Mo ($\gamma$,n) with curve corresponding to $\beta = 0.15$ and $\gamma = 31°$ (blue) as well as $\beta = 0.05$ and $\gamma = 0°$ (cyan) as derived by FRDM [55]. Top: $^{98}$Mo($\gamma$,n) with $\beta = 0.18$ combined to $\gamma = 0°$ (green), $\gamma = 60°$ (cyan), $\gamma = 32°$ (pink, [26]), and $\gamma = 25°$ (blue, our values).

Finally, the good overall agreement between the cross sections from different experiments with the ones calculated by TALYS on the basis of the new E1-parameterization [22] and a reasonable selection of deformation parameters is remarkable. Although in $^{98}$Mo and especially in $^{100}$Mo, the <u>apparent</u> width of the GDR reaches or even exceeds 6 MeV, three overlapping Lorentzians of smaller width, but shifted by the proper amount (see Table II) lead to a very good description of the resonance and to the tail region (*cf.* Fig. 13). The latter is especially sensitive to the width $\Gamma$ and its possible energy dependence. The ansatz allows the use of a GDR spreading width depending smoothly on the energy of the GDR component, and thus only weakly on A and Z. Additionally, it remains in agreement with the TRK sum rule, which is based on very general considerations [57-58], and which we find to be respected at the low excitation energies studied.

This does not exclude a possible violation at higher energies where, *e.g.*, pionic degrees of freedom may come into play. Nevertheless also our data indicate some excess of the 'pygmy' type, which has attracted attention recently – albeit constituting a few % of the E1 strength only.

### C. Apparent excess over a smooth strength distribution

A comparison of the combined data for the Mo-isotopes with the smooth Lorentzian curve from the parameterization used here shows some enhancements of the strength over the extrapolated tail of the GDR. As seen in Fig. 13, the approximations made for M1-strength in RIPL-2 [23] – which are used in the code TALYS [64] – result in a cross-section increase of several % over the E1-contribution. But it is also obvious from the Figure that there is an additional surplus.

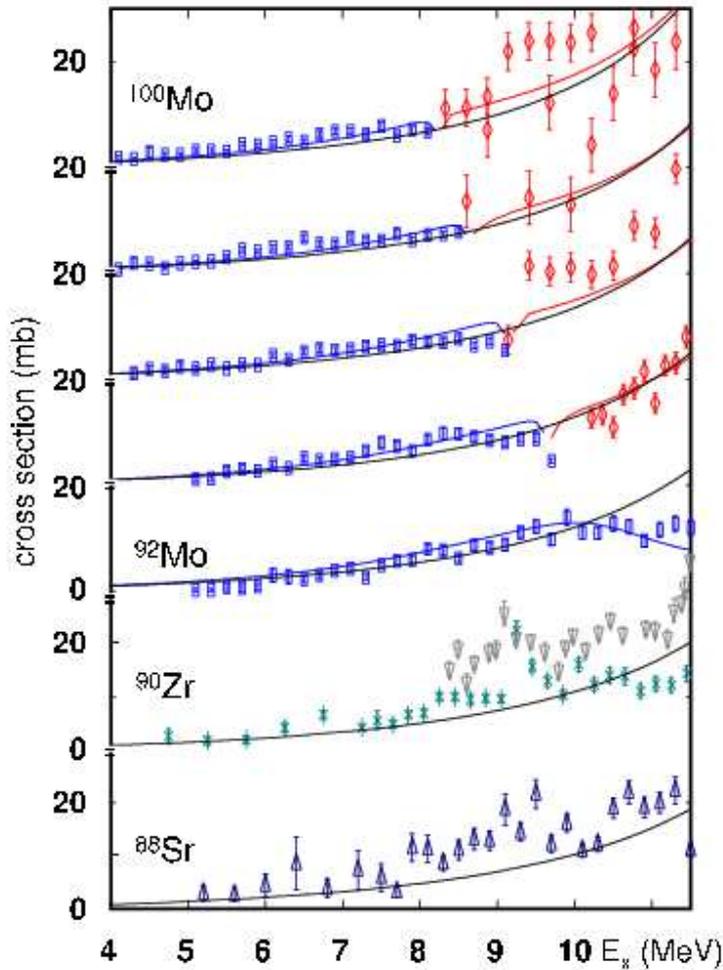

**Fig. 13**: (Color online) Experimental cross sections for photon induced processes in $^{100,98,96,94,92}$Mo (from top). The plot replicates in linear scale the data already shown in Fig. 11 coming from the present scattering experiment (□, blue). They are shown together with the tail of the GDR-Lorentzian (black) as parameterized in [22] and the result of a calculation with the code TALYS [64] using it as input (blue). This code includes M1 strength in addition to E1. For comparison the respective data for targets of $^{88}$Sr

($\Delta$, dark blue) and $^{90}$Zr (**x**, green) have been included at the bottom of the Figure; below 7.5 MeV these data are shown in enlarged bins. Also included here ($\nabla$, grey) is the absorption cross section from an experiment [4] on $^{90}$Zr, which was performed previously with tagged photons, and thus could be analyzed in a quite different manner; the agreement between the two data sets on an absolute scale is satisfactory.

Two flat maxima at about 7.5 and 8.5 MeV may be stated for $^{94}$Mo, whereas in $^{100}$Mo and $^{98}$Mo extra strength appears closer to 7 MeV. As depicted in Fig. 11 the ($\gamma$,n)-data show irregularities just above $S_n$ with large statistical uncertainties. As a basis for a discussion of 'pygmy' structures in Mo as compared to neighboring nuclei Fig. 13 shows – in linear scale – not only the results of the present study, but also the respective data [4, 27] for other even N=50 nuclei. Some excess of the dipole strength above a Lorentzian can be identified for $^{90}$Zr [27] near 6.5 and 9 MeV. Earlier, a cross section enhancement at 9.1 MeV was found in this nucleus and most of that strength was shown not to be of M1 character [68]. For the spherical nucleus $^{88}$Sr [27] and at energies clearly below $S_n$, *i.e.*, at 6-7 and 9-10 MeV, a significant overshoot of the experimental data over the Lorentzian is observed. For N=50 it can thus be stated, that 'pygmy' structures are losing strength with increasing distance from the Z=38 sub-shell, similarly as they disappear more and more in the Mo-isotopes when going away from N=50. 'Pygmy resonances' have been studied extensively for N=82, and they show up as significant structure above the extension of the GDR. This is also true for $^{208}$Pb where one observes strong peaks between 6 and 7 MeV and possibly also at 11 MeV (as in $^{90}$Zr); this may be considered an indication for a sequence of intermediate structures – not just one. A possible explanation of the preferential observation near shell closure may address single-particle configurations not completely 'dissolved' in the GDR.

In all the cases mentioned so far, a group of discrete spectral lines, apparently standing out especially strong from the quasi-continuum, have been identified as 'pygmy' structure. These results, as well as the current Mo data [18], have been related to broad resonance-like "pygmy" structures recently observed in $^{130}$Sn and $^{132}$Sn [7] at an energy just above 9 MeV. This was done, albeit the neutron excess in these Sn isotopes is much higher, possibly favoring collective modes hitherto unknown. In a very recent study [69] on $^{117}$Sn($^3$He,$^3$He) an enhancement at about 8 MeV is seemingly spread out in energy, and thus resembles a Lorentzian shaped resonance. It is especially prominent in comparison to RPA results [45], or to a strength function with photon energy dependent width, both documented in RIPL2 [23]. As shown in Figs. 7 and 9 of the present paper, these two approaches are not only failing to describe our new low energy Mo-data, the RPA calculation [45] also misses the exact shape of the GDR. Using a tagged photon beam and a $^{nat}$Sn target, a remarkable enhancement in the photo-absorption cross section near 8 MeV – of more than a factor of two as compared to a Lorentzian extrapolation – was observed [4] many years ago.

More conclusive statements about the nature of the excess strength at these sub-GDR energies require a more extensive study of the A and Z-dependence. A detailed study of their statistical properties in the sense of RMT may give further insight as well as information coming from investigations of the same excitation region using different incident channels [8]. Especially it has to be determined, if the 'pygmy resonances' indicate extra absorption in the entrance channel or rather a structure in the final nucleus' level density. Here one should consider how a non-continuous level density, which already seemed to be the cause of the structure seen [67] in $^{92}$Mo($\gamma$,p), would modify the cross section calculations of Hauser-Feshbach type. Often the level density is not determined experimentally, as the experiments are performed with low energy resolution, not allowing the observation of sufficiently many discrete levels. In view of the fact that the pygmy structures are best seen after an energy average, our approach of defining a procedure to accurately predict the GDR tail is helpful for a quantification of the eventual excess.

As the feeding corrections become too large to allow any statement about strength below 4 MeV. the present study is limited to energies above. In our previous paper on $^{98}$Mo and $^{100}$Mo it was shown [19, Fig.15] that data taken previously at low endpoint energy reveal an integrated E1 strength continuously decreasing, when $E_x$ falls below 4 MeV. This is at variance with a photon strength function $f_1(E)$ obtained from $^3$He induced reactions [13] showing a strong peak near 1 MeV. A recent paper [70] on $^{95}$Mo(n,$\gamma$)$^{96}$Mo did also rule out such a non-monotonous $f_1(E)$, but the strength applied in the analysis of these data is not in full accord to our findings: At an energy of 6 MeV our $f_1(E)$ is larger by nearly a factor of 4 as compared to the one used in ref. 70, but also the average level distance $D$ used there is at least by a factor of 3 larger than our estimate, obtained from the considerations discussed in connection to Fig. 6. Experiments only observing electromagnetic decay without a direct measurement of the decay width [13, 69, 70] are usually mainly sensitive to $f_1(E)/D$, whereas our study involving also electromagnetic excitation can deliver direct information on $f_1(E)$ alone – as was outlined in section III.A. As in the abovementioned paper [70] the statement is made: *"It should be emphasized that . . we can*

*easily test the acceptability of different models of $f_1(E)$ and level density, but it is not possible to determine a best fit"* a correlated study of $(\gamma,\gamma)$, $(\gamma,n)$ and $(n,\gamma)$ for the same final nuclei is desirable.

## VI. CONCLUSIONS

The response of nuclei to dipole radiation can well be studied at a bremsstrahlung facility like ELBE. Using a sufficiently high endpoint energy and correcting photon scattering data for non-resonant and inelastic fractions allowed us to combine the dipole strength functions $f_1$ obtained from nuclear resonance fluorescence with existing nuclear photo effect measurements. Using activation techniques we reinvestigated some of the latter data, and normalized them accurately. The good match between the data for above and below particle thresholds is remarkable, and together they span a the energy region from low excitation up to the GDR. Our approach allows the extraction of the dipole strength for a wide range, and our results do not depend on *a priori* level density information. For the Mo data discussed here the level distances $D$ approach the experimental resolution, and thus the spectra studied are dominated by Porter-Thomas fluctuations in most of the energy range. Consequently the chaotic structure in the photon scattering excitation functions – with a bremsstrahlung beam they are observed simultaneously over a wide range – has to be accounted for. A proper extraction of strength information cannot ignore the fluctuating quasi-continuous part of the experimental yield, and thus an averaging procedure over an interval $\Delta \geq D$ was introduced to extract the dipole strength function $f_1(E)$. The contribution from the unavoidable non-resonant scattering was simulated and subtracted. Complementing the nrf studies by photo-activation experiments allowed to assess the weak influence of different optical model parameters on the results of Hauser-Feshbach calculations. More important is that the activation data verify a previous conjecture [11, 12] about the need of a renormalization of existing photo-neutron data [10, 20]. This point gives further support to our finding that in the Mo-isotopes studied here, the TRK sum-rule is considerably better satisfied than inferred from data before.

A comparison of $f_1(E)$ to a Lorentzian extrapolated from the GDR with a deformation induced splitting, properly adjusted to the ground state shape parameters, shows a very good agreement with the data below and above the particle separation energies. Two important ingredients of the parameterization used are the direct account of deformation, and a photon-energy independent width. The first point was especially well studied in the even Mo-isotopes due to their wide variation in axial and triaxial deformation. For the second finding, the wide energy range available for a comparison to experimental data played an important role. The proposed procedure of an explicit inclusion of the widening of the GDR due to deformation allows a clear-cut separation from the width, caused by the spreading into the quasi-continuum underlying it. Combined to the fixing of the absolute height by the sum rule [57, 58], and by allowing for triaxiality, we achieved a very reliable extraction of the GDR parameters. Especially the spreading width is defined more accurately than was accomplished by just fitting the GDR near its maximum [10, 23, 24]. The variation of the width with the GDR energy, and thus also with the nuclear mass number A, turned out to be surprisingly smooth.

The GDR shape is well described in size and energy dependence by the new Lorentzian parameterization, which already before had been shown [22] to be valid globally for all nuclei with A>80, for which respective data exist. In nuclei with 'pygmy' structures on the low energy slope of the GDR this expression is valuable as a well quantified reference. Such structures, often reported for nuclei near magic shells, become weaker in the Mo-isotopes with the neutron number increasing from N=50. The pygmy structures also fade away when going from $^{88}$Sr and $^{90}$Zr to $^{92}$Mo, *i.e.* by leaving the proton sub-shell closure. Going away from shells leads not only to an increased level density, but also to appreciable ground state deformation, which enters crucially in our parameterization for the GDR shape. The dipole strength near $S_n$ and $S_p$ was shown to be enhanced by the lowering of the energy of one of the GDR components with deformation. The photon strength there is of technical importance for transmutation processes based on radiative neutron capture, as well as for the understanding [1,2] of various processes participating in the cosmic nucleosynthesis. This is the reason for the importance of the proposed [22] parameterization for predictions on the dipole strength at the particle emission thresholds. The hope is that it is of use also when no GDR-data are available, which is the case for nearly all nuclei outside the valley of stability. Finally, it should be stated, that many of the prescriptions contained in the RIPL-2 database [23], prepared at IAEA and widely used for the evaluation of nuclear processes, are at variance with the data presented here as well as with the Lorentzian parameterization shown to describe them.

## Acknowledgements

P. Michel and the ELBE-Crew made these experiments possible by delivering optimum beams. Help from and discussions with F. Becvár, T. Belgya, D. Bemmerer, P. Crespo, M. Fauth, B. Kämpfer, N. Nankov, M. Krticka, R. Roth, G. Rusev, K.D. Schilling, E. Trompler and S.Q. Zhang are gratefully acknowledged. This work was funded by DFG under the contract Do466/2 and by the 6$^{th}$ EU-Framework Program under 036434; support also came from BMBF and GSI Darmstadt.